# RF gymnastics in synchrotrons

*R. Garoby*
CERN, Geneva, Switzerland

**Abstract**
The RF systems installed in synchrotrons can be used to change the longitudinal beam characteristics. 'RF gymnastics' designates manipulations of the RF parameters aimed at providing such non-trivial changes. Some keep the number of bunches constant while changing bunch length, energy spread, emittance, or distance between bunches. Others are used to change the number of bunches. After recalling the basics of longitudinal beam dynamics in a hadron synchrotron, this paper deals with the most commonly used gymnastics. Their principle is described as well as their performance and limitations.

## 1 Introduction

RF systems in synchrotrons are primarily specified for beam acceleration in variable energy machines or for bunching in accumulators. At a later stage of the design, and quite often after the machine is built, the need to tailor further the longitudinal beam characteristics like bunch length, energy spread, distance between bunches, number of bunches etc., frequently occurs. 'RF gymnastics' involving the modulation of the RF parameters are then considered to help obtain the required performance [1].

As the high-energy frontier gets higher and higher, the cost of an accelerator complex increases accordingly, as well as the interest in gymnastics which give the possibility to adapt such a facility for purposes which were not originally foreseen.

## 2 Longitudinal beam dynamics

### 2.1 Conventions

Synchrotron radiation will not be considered so that the following analysis is relevant only for hadrons.

The longitudinal phase plane has time (or phase) as x axis and energy (or momentum) as y axis. The following variables characterize a particle:

– Charge: $q$

– Rest energy, energy: $E_0$, $E$

– Speed, momentum: $v$, $p$

– Relativistic parameters: $\gamma$ ($\gamma = E/E_0$), $\beta$ ($\beta = v/c$)

– Revolution period in the synchrotron: $T$.

The synchrotron parameters are the following:

– Momentum compaction factor, transition gamma: $\alpha_P$, $\gamma_T$

– Parameters of the synchronous particle: $E_S$, $v_S$, $p_S$, $\gamma_S$, $\beta_S$, $T_S$. The synchronous particle is defined as the particle whose energy $E_S$ and phase $\phi_S$ (measured with respect to the zero crossing with positive slope of the sinusoidal RF waveform at the lowest harmonic ($h_1$)), are such that it sees the same accelerating voltage over successive turns in the accelerator.

The total voltage *V(t)* results from the contributions of RF systems with voltages $V_1(t)$, $V_2(t)$,…

$$V(t) = \sum_{i=1}^{n} V_i(t) \tag{1}$$

If resonant structures are used, the voltage functions are sine-waves with $h_i$ periods per revolution and a relative phase $\theta_i$.

$$V_i(t) = \hat{V}_i \sin(h_i \omega_R t + \theta_i), \quad \text{with} \quad \omega_R = \frac{2\pi}{T_S}. \tag{2}$$

## 2.2 Motion in the longitudinal phase plane

### 2.2.1 Equations of motion

The motion of particles is analysed in the frame of the synchronous particle. The x coordinate is the phase difference $\Delta\phi = \phi - \phi_S$ measured at the lowest harmonic ($h_1$), and the energy coordinate is $\Delta E = E - E_S$ (or $\Delta p = p - p_S$). The tracked and the synchronous particles having different revolution periods, the phase difference $\Delta\phi$ changes at every revolution according to Eq. (3):

$$d\Delta\varphi = 2\pi h_1 \frac{(T - T_S)}{T_S} = 2\pi h_1 \frac{\Delta T}{T_S}. \tag{3}$$

The rate of change of the phase is then

$$\frac{d\Delta\varphi}{dt} = \frac{2\pi h_1}{T_S} \frac{\Delta T}{T_S}. \tag{4}$$

In a synchrotron the relative difference in revolution period is proportional to the relative difference in momentum or energy:

$$\frac{\Delta T}{T_S} = \eta \frac{\Delta p}{p_S} = \frac{\eta}{\beta^2} \frac{\Delta E}{E_S}, \tag{5}$$

where

$$\eta = \frac{1}{\gamma_T^2} - \frac{1}{\gamma^2}.$$

From Eqs. (4) and (5), the x component of the particle speed is given by

$$\frac{d\Delta\varphi}{dt} = 2\pi h_1 \eta \frac{1}{\beta^2 T_S} \frac{\Delta E}{E_S}. \tag{6}$$

The *y* component of the particle speed is the rate of change of its energy with respect to the synchronous particle and is given by Eq. (7):

$$\frac{d\Delta E}{dt} = \frac{q}{T_S}\left[V(\Delta\varphi + \varphi_S) - V(\varphi_S)\right]. \tag{7}$$

### 2.2.2 Case of a single RF harmonic

When a single RF system is used, the voltage can be expressed as

$$V(\phi) = \hat{V} \sin\phi \tag{8}$$

and Eq. (7) simplifies into

$$\frac{d\Delta E}{dt} = \frac{q}{T_S} \hat{V}\left[\sin(\Delta\varphi + \varphi_S) - \sin\varphi_S\right]. \tag{9}$$

The motion described by Eqs. (6) and (9) has the following first integral characterizing closed trajectories of particles oscillating around the synchronous one:

$$\frac{1}{2}\left(\frac{d\Delta\varphi}{dt}\right)^2 + \frac{2\pi h_1 \eta q \hat{V}}{\beta^2 T_S^2 E_S}\left[\cos(\Delta\varphi + \varphi_S) + \Delta\varphi \sin\varphi_S\right] = \text{constant} . \tag{10}$$

There is a limit to the amplitude of these oscillations. The corresponding trajectory is called the separatrix, and the enclosed region is the bucket whose area is the acceptance. The separatrix crosses the phase axis at the extreme phase elongation:

$$\Delta\varphi_{EXT1} = \pi - 2\varphi_S . \tag{11}$$

The other extreme phase elongation is the solution of

$$\cos(\Delta\varphi_{EXT2} + \varphi_S) + \Delta\varphi_{EXT2}\sin\varphi_S = -\cos\varphi_S + (\pi - 2\varphi_S)\sin\varphi_S . \tag{12}$$

The extreme excursion in energy is obtained when $\Delta\phi = 0$ rad.:

$$\Delta E_{MAX} = \sqrt{\frac{E_S \beta^2}{\pi h_1 \eta} q \hat{V}\left[(\pi - 2\varphi_S)\sin\varphi_S - 2\cos\varphi_S\right]} . \tag{13}$$

Figure 1 illustrates the case of a stationary bucket (constant $B$ field in the main dipoles and no acceleration of the synchronous particle) below transition energy ($\phi_S = 0$ rad.). The separatrix extends from $-\pi$ to $+\pi$ radians. The speed of a moving particle inside the bucket is shown. If it is an extreme particle of a stable population, its trajectory is the contour enclosing all others. This set of particles is called a bunch and the area inside the contour is its emittance.

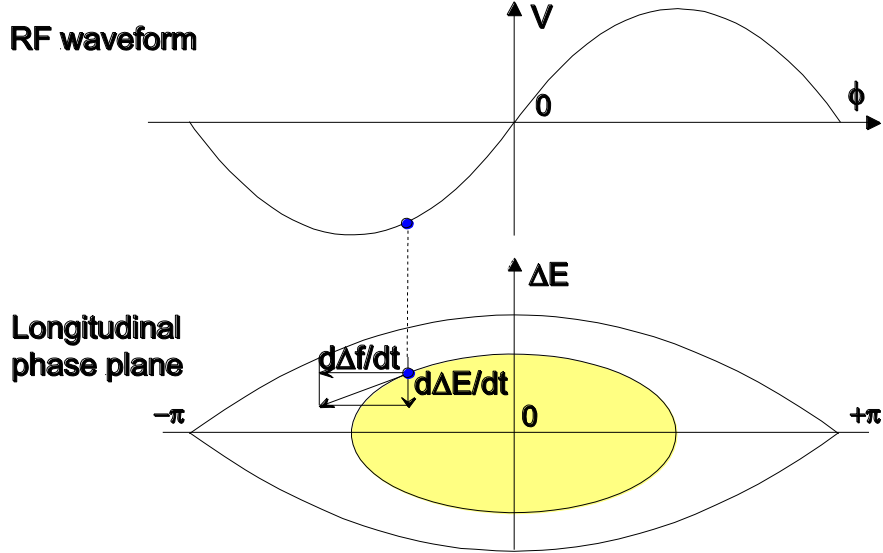

**Fig. 1:** Trajectories in a stationary bucket

For small amplitude of oscillation, Eqs. (6) and (9) represent a simple harmonic oscillator at the synchrotron frequency $\omega_S$:

$$\omega_S = \sqrt{\frac{2\pi h |\eta| q \hat{V}\cos\varphi_S}{\beta^2 T_S^2 E_S}} . \tag{14}$$

At constant emittance, the peak excursions in phase and energy scale like

$$\Delta\hat{\varphi} \propto k \quad \text{and} \quad \Delta\hat{E} \propto \frac{1}{k} \quad \text{with} \quad k = \left[\frac{|\eta|}{E_S q \hat{V}\cos\varphi_S}\right]^{\frac{1}{4}} . \tag{15}$$

## 2.3 Effect of changing RF parameters

### 2.3.1 Adiabaticity

If the RF parameters are changed at a slow rate with respect to the smallest frequency of oscillation of the particles in the bunch, the distribution of particles is continuously at equilibrium and depends only upon the instantaneous value of these parameters. Such an evolution is called 'adiabatic'. The degree of adiabaticity is assessed with the adiabaticity parameter [2] defined as

$$\varepsilon = \frac{1}{\omega_S^2}\left|\frac{d\omega_S}{dt}\right|. \tag{16}$$

A process is typically considered adiabatic when $\varepsilon < 0.1$.

### 2.3.2 Liouville's theorem

The longitudinal motion that we consider is conservative (i.e., there is no energy dissipation effect like synchrotron radiation). Liouville's theorem which states that the local density of particles in the longitudinal phase plane is always constant [3], is then applicable. An implicit consequence is that any RF gymnastics is in principle reversible.

When an adiabatic process is used, this helps determine the particle distribution (or bunch shape) in the final state without having to take into account the intermediate ones (Fig. 2). The area occupied by particles ('emittance') is constant and always limited by a stable trajectory.

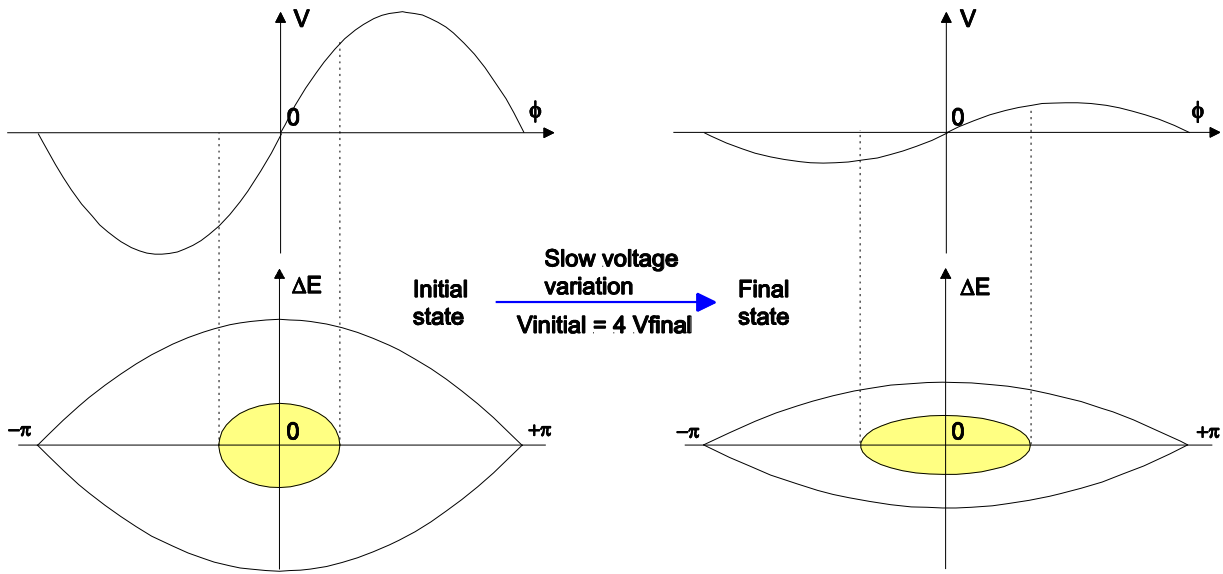

**Fig. 2:** Adiabatic RF voltage reduction

When a non-adiabatic gymnastic is applied, the consequences are less obvious and a detailed tracking is required to evaluate the final particle distribution (Fig. 3). Although the area occupied by particles is also constant, its contour is usually not a stable trajectory in the final state. The final emittance generally has to be considered as increased to the value of the smallest area limited by a stable trajectory which contains all particles ('macroscopic' emittance).

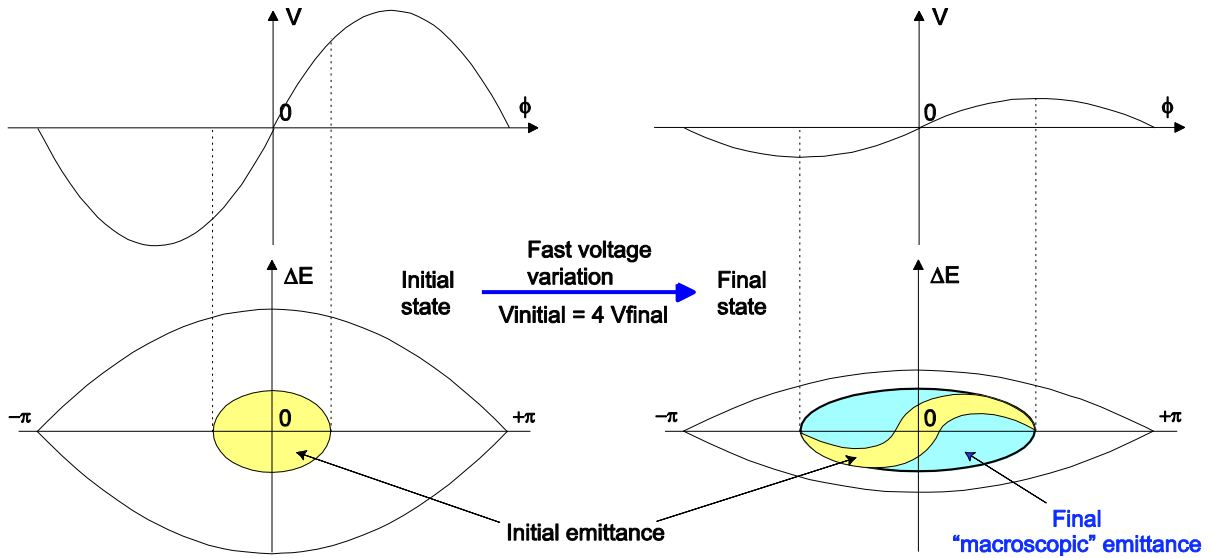

**Fig. 3:** Non-adiabatic RF voltage reduction

## 3 Single-bunch gymnastics

### 3.1 Bunch compression

To preserve the longitudinal emittance and guarantee reproducible beam performance, the contour of the bunches entering a synchrotron must correspond to stable trajectories in the longitudinal phase plane. Such a condition is called 'longitudinal matching'. This often requires changing the ratio bunch length/energy spread of the bunches in the previous machine, and generally bunches must be made shorter. When adiabatic variation of the RF voltage cannot be used to provide the proper beam characteristics, non-adiabatic processes are applied. The corresponding gymnastics are called 'bunch compression', 'bunch rotation' or even 'phase rotation' [4, 5].

The principle (Fig. 4) is to let a bunch, initially elongated in phase, rotate in a maximum height bucket, and to eject it when it is shortest. Even with a single RF system, various techniques can be used for stretching the bunch:

– Reducing adiabatically the RF voltage $V$, the bunch length increases in proportion to $V^{-1/4}$ [Eq. (15)]. This technique has the drawback of requiring a very large dynamic range in $V$ and of becoming very slow to remain adiabatic at low voltages [see Eqs. (14) and (16)].

– Reducing abruptly the RF voltage, a bunch rotation is triggered which provides, after a quarter of a turn in the phase plane, a bunch length proportional to $V^{-1/2}$. This process is faster and more efficient than the previous one but is more demanding for the transient response of the cavity and the beam servo-loops.

– Switching by $\pi$ radians the phase of the RF, the bunch becomes centred on the unstable phase and stretches quickly along the separatrix. This technique is also fast and does not in principle require any voltage change, but it needs fast response of the RF system. The fact that the resulting bunch is tilted with respect to the phase axis implies that it will suffer more from non-linearities when rotating in the phase plane for compression.

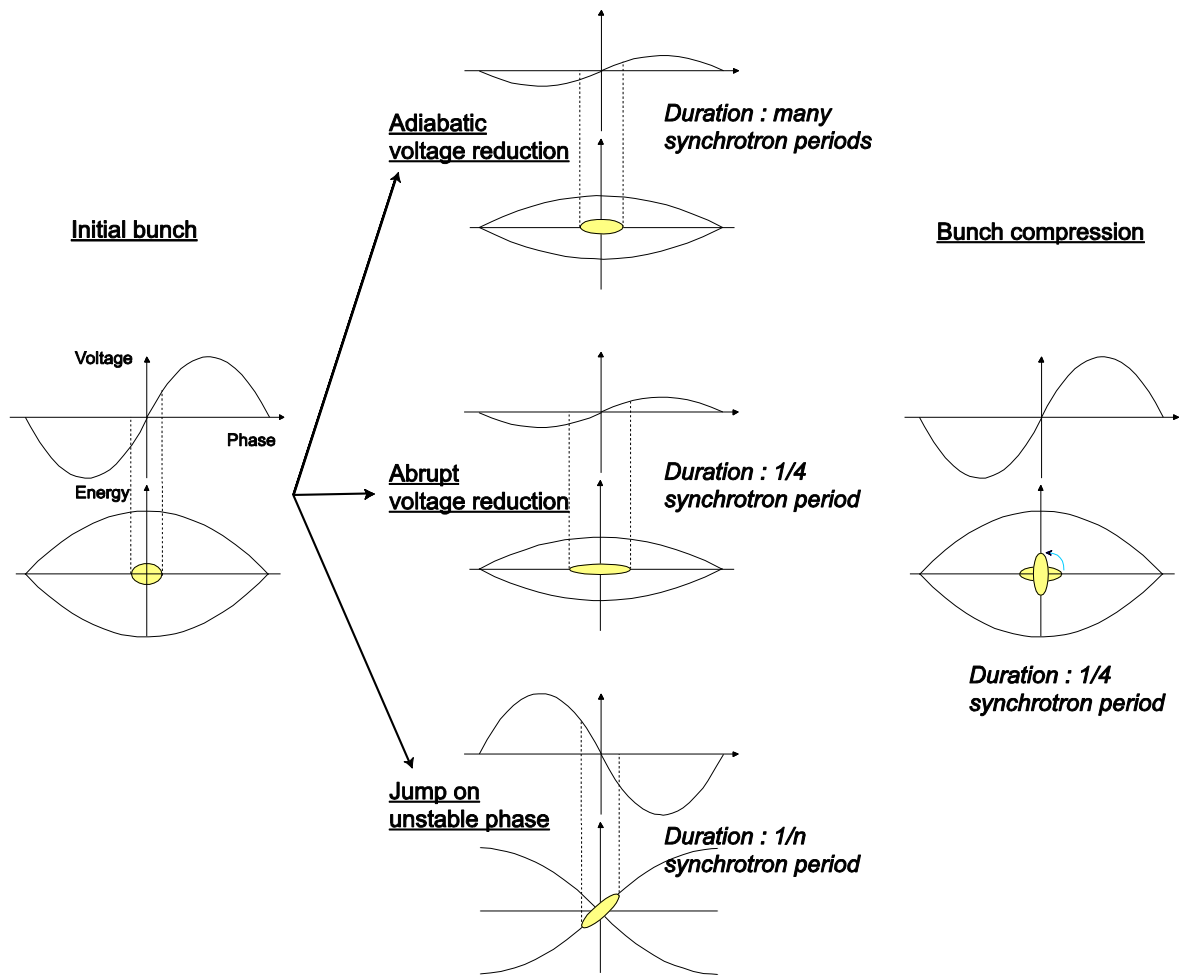

**Fig. 4:** Bunch compression

The quality of the compression process for an elongated perfectly 'straight' bunch depends upon its length and normalized emittance (ratio between emittance and acceptance). This is illustrated in Fig. 5 which shows the bunch at the beginning and at the end of rotation. An initially extreme particle along the energy axis (B0) becomes extreme in phase after rotation (B1) under the effect of a quasi-linear focusing voltage approximated by the slope at zero phase of the RF sine-wave. On the contrary, an initially extreme particle along the phase axis (A0) experiences a non-linear and on average smaller focusing voltage during rotation, which results in a slower motion. In the time it takes for B0 to move to B1, A0 only moves to A1.

For a given normalized emittance, the minimum bunch length is obtained approximately when A1 and B1 are at the same phase. This defines an optimum initial bunch elongation which is represented in Fig. 6. This figure also gives the minimum length achieved after rotation and the equilibrium length of a bunch of the same emittance in the rotation bucket. A compression efficiency can be defined as the ratio between that equilibrium bunch length and the length after rotation in optimum conditions. This efficiency is also shown in Fig. 6.

As a typical example, a bunch filling 1% of the bucket has

- an adiabatic bunch length of ~0.073 bucket length,
- an optimum length before rotation of ~0.27 bucket length,
- and a minimum bunch length of ~0.02 bucket length corresponding to a compression efficiency of ~3.7.

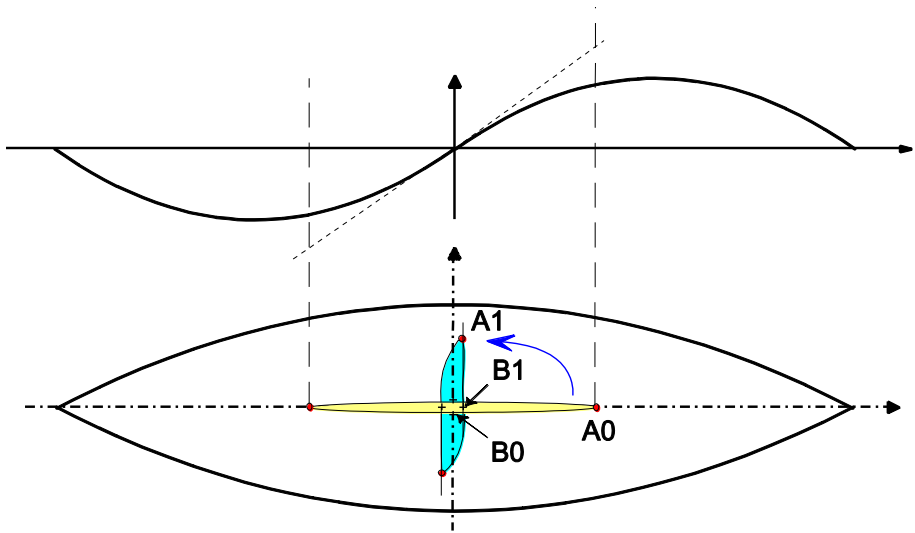

**Fig. 5:** Optimum bunch rotation

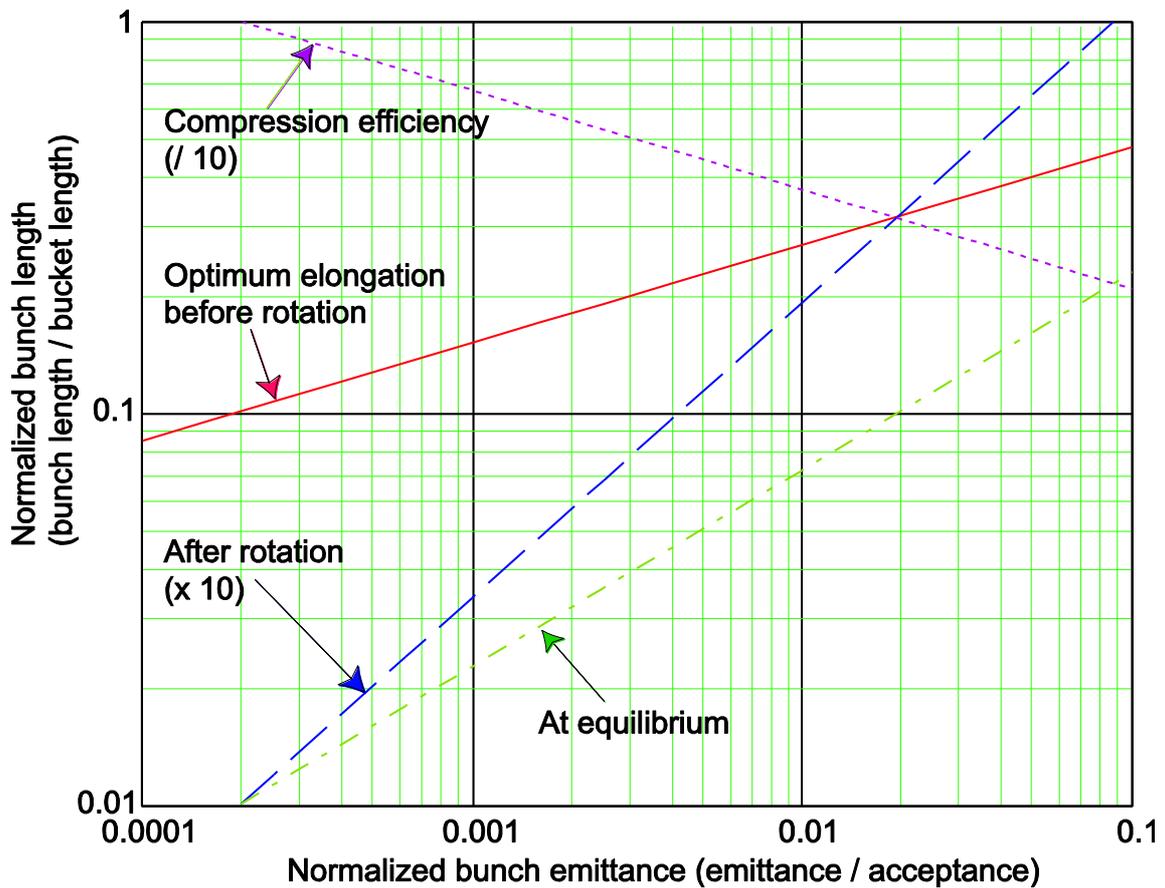

**Fig. 6:** Bunch rotation parameters

Higher compression ratios/higher compression efficiencies can be obtained using more complicated gymnastics involving multiple RF harmonics and/or phase and amplitude modulations.

### 3.2 Longitudinal controlled blow-up

Blow-up techniques have been developed to help stabilize high-intensity beams by increasing the 'macroscopic' emittance in a controlled way while providing an adequate distribution of particles with sharp edges and no tails. A typical and commonly-used technique is based on the superposition of a phase-modulated high frequency ($V_H$, $h_H$) to the RF normally holding the beam ($V_1$, $h_1 \ll h_H$) [6, 7].

The high-frequency phase-modulated voltage can be expressed as

$$V_H = \hat{V}_H \sin(h_H \omega_R t + \alpha \sin \omega_M t + \vartheta_H), \quad (17)$$

$\alpha$ being the peak phase modulation, $\omega_R$ the modulation frequency, and $\theta_H$ a phase constant.

This acts as a perturbation to the motion of particles in the bucket of the main RF system. Resonances can be induced which create a re-distribution of density in the bunch. Large non-linearities in the motion accelerate filamentation and contribute to the fast disappearance of the density modulations induced by the high-frequency carrier. Among the different distributions that can be obtained, parabolic ones are generally preferred.

The blow-up parameters are in practice optimized either on the real accelerator or using computer simulations. Typical ranges of values applied in such cases are shown in Table 1.

A slower but still well-controlled blow-up can also be attained with a smaller harmonic ratio. This is especially valuable in slow cycling synchrotrons.

**Table 1:** Typical blow-up parameters

| | $\dfrac{\hat{V}_H}{\hat{V}_1}$ | $\dfrac{h_H}{h_1}$ | $\alpha$ (rad) | $\dfrac{\omega_M}{\omega_S}$ | Duration |
|---|---|---|---|---|---|
| Typical range | 0.1 to 0.3 | > 10 for fast blow-up | $0.8\pi$ to $1.2\pi$ | 3 to 12 | $\geq 20 \dfrac{2\pi}{\omega_S}$ |

## 4 Multi-bunch gymnastics

### 4.1 Debunching–rebunching

Debunching–rebunching is the most conventional way to change the number of bunches [5, 8]. It has to take place at constant energy and hence at constant field in the main bending dipoles because of the absence of RF for a significant period of time. At the end of debunching the beam is continuous and ideally without any azimuthal modulation of the linear density of particles. Rebunching is the reverse process during which a different RF harmonic number is used, and the beam progressively gets an azimuthal modulation of density and is finally fully bunched on the new harmonic.

Iso-adiabatic debunching is generally used to minimize longitudinal emittance blow-up. The reduction of the RF voltage from $V_{I\_deb}$ to $V_{F\_deb}$ is done at constant adiabaticity [see Eq. (16)]:

$$V(t) = \dfrac{V_{I\_deb}}{\left[1 - \left(1 - \sqrt{\dfrac{V_{I\_deb}}{V_{F\_deb}}}\right)\dfrac{t}{t_R}\right]^2} \quad (18)$$

where $t_R$ is the moment of suppression of the RF voltage after reaching the minimum controllable level $V_{F\_deb}$. This is illustrated in Fig. 7. The process takes more time when $V_{F\_deb}$ is made smaller:

$$t_R \approx \dfrac{1}{\omega_S(V_{F\_deb})} \propto \dfrac{1}{\sqrt{V_{F\_deb}}}. \quad (19)$$

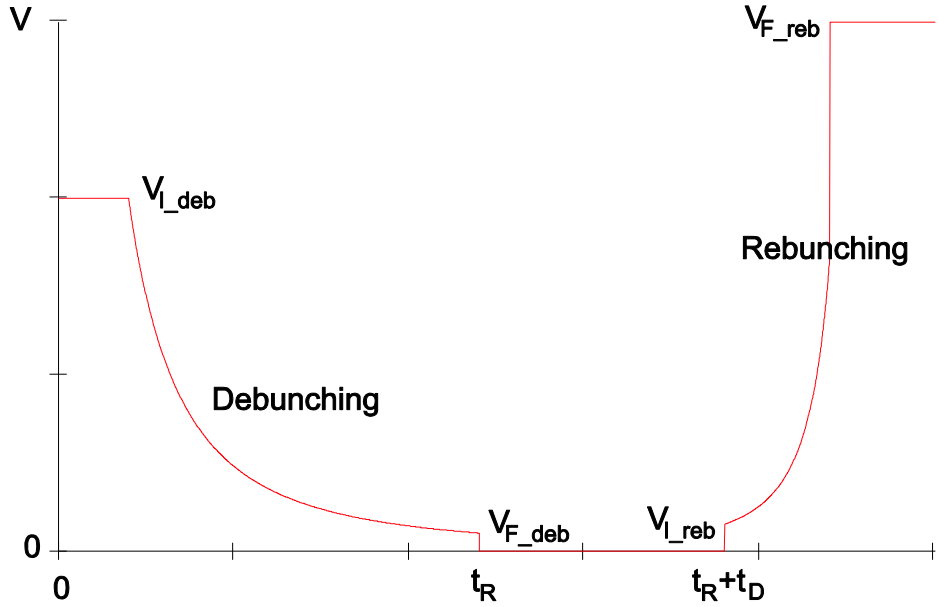

**Fig. 7:** Voltages for iso-adiabatic debunching–rebunching

During this voltage reduction, the bunch progressively lengthens [proportionally to $V^{-1/4}$ at the beginning according to Eq. (15)]. Under the voltage $V_{F\_deb}$ the beam is generally still bunched and some time $t_D$ is required without voltage for the particles to drift in azimuth and for debunching to be obtained. This results in a blow-up of the macroscopic emittance which depends upon the normalized bunch emittance in the final bucket as shown in Fig. 8. In the typical case where the bunch finally fills the bucket completely, the emittance is multiplied by $\pi/2$.

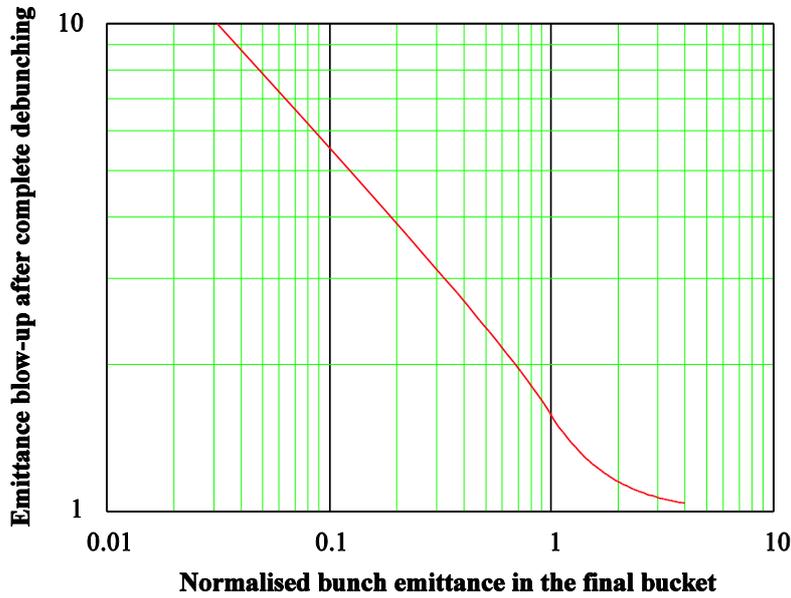

**Fig. 8:** Emittance blow-up after iso-adiabatic debunching

A reference debunching time can be defined as the time taken for the particles of successive bunches to begin to overlap in azimuth:

$$t_{D\_classic} = \frac{\pi - \Delta\varphi}{h\omega_R |\eta| \frac{\Delta p}{p}}, \qquad (20)$$

where $\Delta\phi$ and $\Delta p$ are the full spreads in phase and momentum of the bunch under $V_{F\_deb}$.

A good-quality debunching with a small residual density modulation requires $t_D \gg t_{D\_classic}$.

Iso-adiabatic rebunching is generally used after debunching is completed. It is a time-reversed version of iso-adiabatic debunching, starting abruptly at the level $V_{I\_reb}$ and rising progressively to $V_{F\_reb}$. Similar formulae apply.

### 4.2 Splitting (merging)

Splitting is used to multiply the number of bunches by 2 or 3 and merging is the reverse process [9, 10]. Although limited in use to circumstances where such ratios are of interest, these processes have the remarkable advantage with respect to iso-adiabatic debunching–rebunching of being capable of being quasi-adiabatic and preserving emittance.

Splitting bunches into two is obtained using simultaneously two RF systems with an harmonic ratio of 2. The bunch is initially held by the first system ($V_1$, $h_1$) while the second ($V_2$, $h_2 = 2\,h_1$) is stopped. The unstable phase on the second harmonic is centred on the bunch. As the voltage $V_2$ is slowly increased and $V_1$ decreased, the bunch lengthens and progressively splits into two as illustrated in Fig. 9.

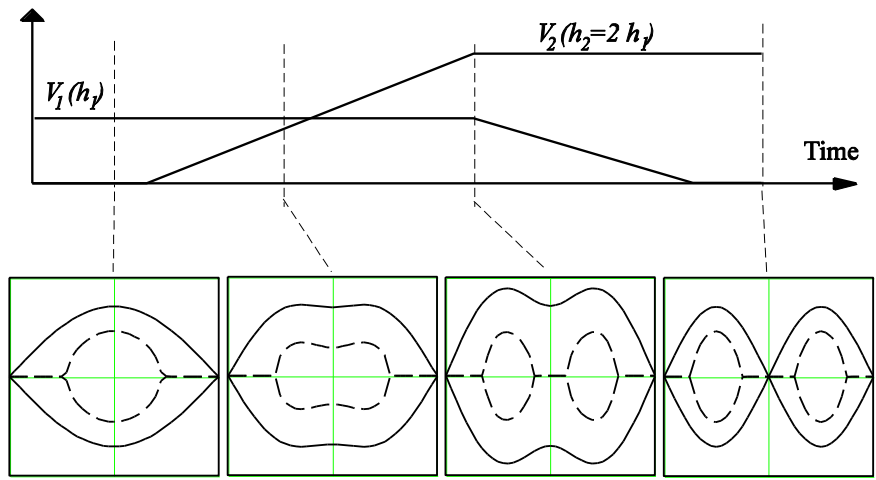

**Fig. 9:** Bunch splitting into two

Good results are consistently obtained when the voltage $V_1(h_1) = V_{1\_sep}$ is such that, at the moment when two separate bunches have just formed, the initial bunch would fill ~1/3 of the bucket acceptance in the absence of second harmonic ($V_2(h_2) = 0$ kV). Voltage variations are generally linear functions of time with a total duration larger than 5 synchrotron periods in the bucket ($V_{1\_sep}$, $h_1$). Each final bunch has ½ the emittance of the initial one, and almost no blow-up is observed.

An illustration of an operational implementation of double splitting in the CERN PS is shown in Fig. 10. A bunch on h = 8 is split into two on h = 16 within 25 ms and no blow-up can be noticed. On the left side of the same figure, the evolution of particle density in the longitudinal phase plane during the process is reconstructed using longitudinal tomography [11].

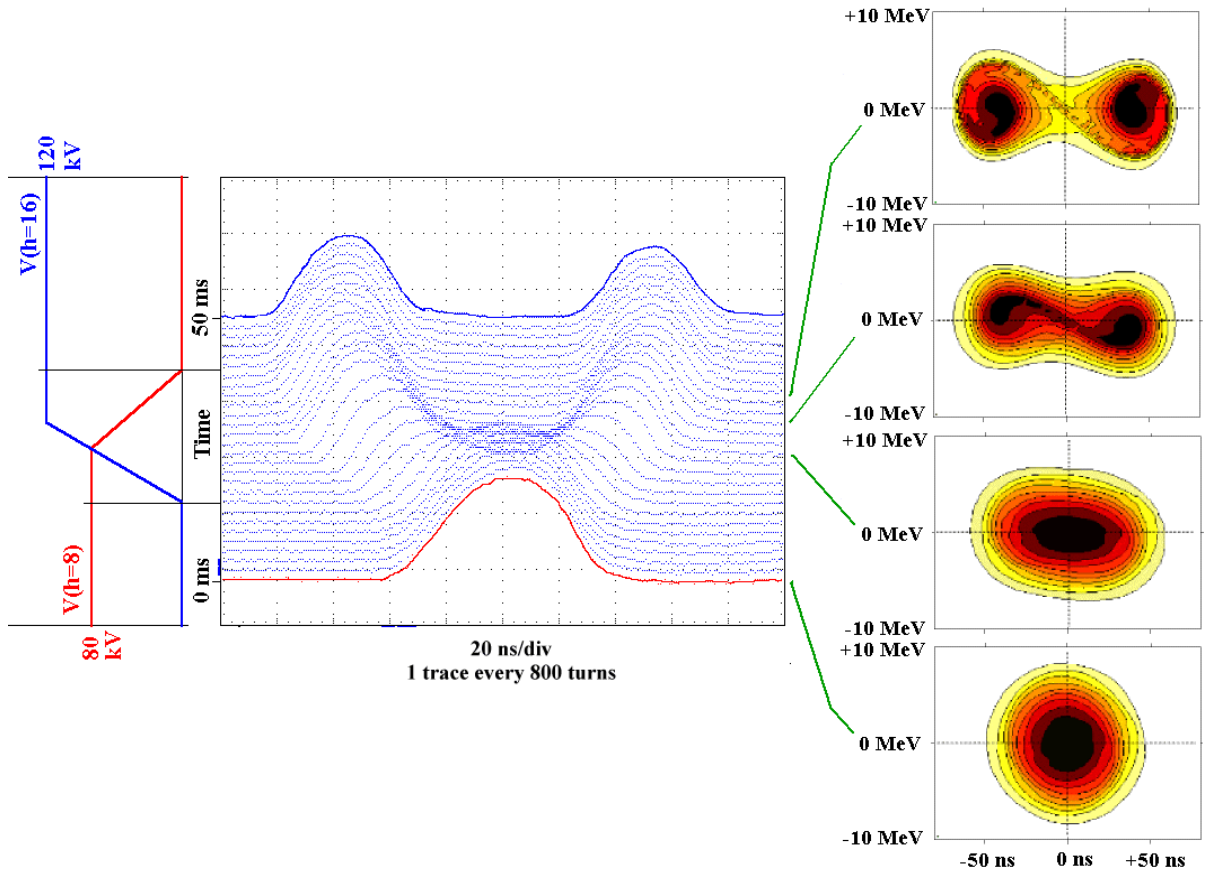

**Fig. 10:** Example of bunch double-splitting from h = 8 to h = 16 in the CERN PS at 3.57 GeV/*c*

Splitting bunches into three requires using three simultaneous RF systems. The relative phases between harmonics as well as the voltage ratios must be precisely controlled for the particles to split evenly into the new bunches and longitudinal emittance preserved. Results as good as for bunch double-splitting have been achieved, and final bunches are 1/3 the emittance of the original one. The voltages and the evolution in longitudinal phase space as a function of time are illustrated in Fig. 11.

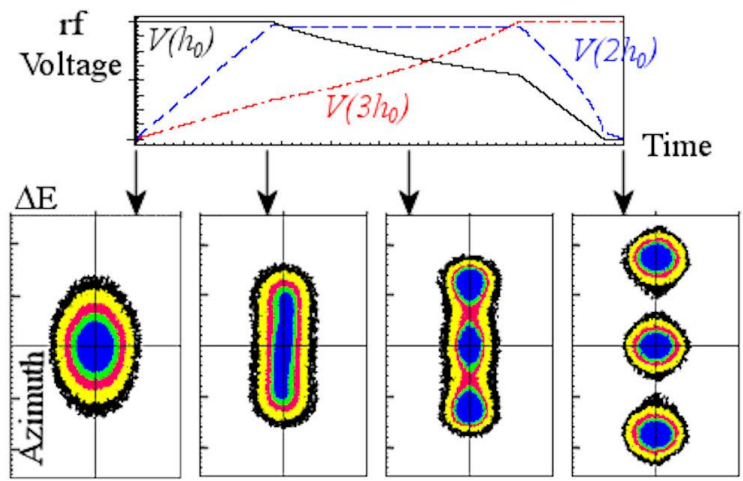

**Fig. 11:** Bunch triple-splitting

### 4.3 Batch compression

Batch compression is a process which keeps the number of bunches constant while concentrating them in a reduced fraction of the accelerator circumference [12]. When exercised at a slow enough rate it can be adiabatic and consequently preserve the longitudinal emittance.

The principle is slowly to increase the harmonic number of the RF controlling the beam as shown in Fig. 12. Starting from harmonic $h_0$, voltage is progressively increased on harmonic $h_1 > h_0$ and decreased to 0 V on $h_0$, so that harmonic $h_1$ finally holds the bunches. The phase on $h_1$ with respect to $h_0$ must be such that the bunches converge symmetrically towards the centre of the batch.

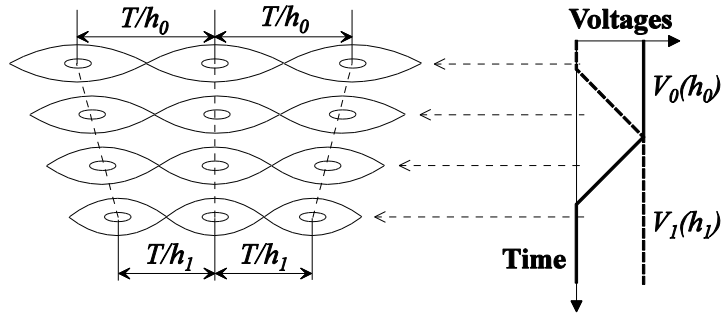

**Fig. 12:** Batch compression

The amount of compression achievable in a single step is limited by the need to maintain a large enough acceptance for the buckets holding the edge bunches. A consequence is that large compression factors are obtained only after multiple batch compression steps, and complicated manipulations of RF parameters are involved. A typical application is given in Fig. 13, where four bunches on h = 8 are finally brought into four adjacent buckets on h = 20: three groups of RF cavities are used which help sweep progressively the harmonic seen by the beam from 8 to 20 in steps of 2 units.

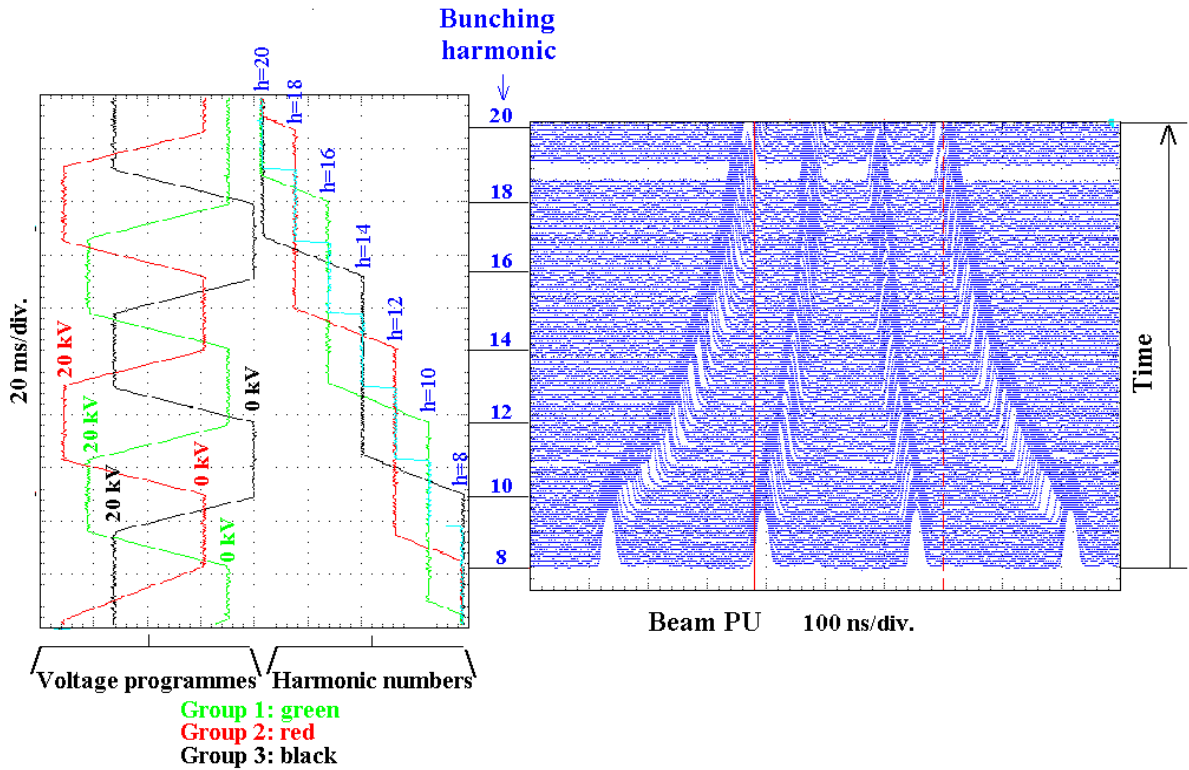

**Fig. 13:** Example of batch compression from h = 8 to h = 20 in the CERN PS at 26 GeV

## 4.4 Slip stacking

Slip stacking is used to superimpose two sets of bunches and double the bunch population [13, 14]. It is non-adiabatic and leads to large emittance blow-ups.

The principle is sketched in Fig. 14. Two different RF frequencies are simultaneously applied. If their difference is large enough ($\Delta f > 2f_s$, where $f_s$ is the synchrotron frequency in the centre of an unperturbed bucket of one family), two families of buckets coexist which drift towards each other because of their frequency difference. Consequently, and provided the acceptance of these buckets ($f = h_0 f_{REV} \pm \Delta f$; $V_{drift}$) is large enough (acceptance > 2 × emittance), the bunches drift with them and tend to slip past each other. When they are superimposed in azimuth, pairs of bunches can be captured in large buckets centred at the middle frequency ($f = h_0 f_{REV}$; $V_{capture}$).

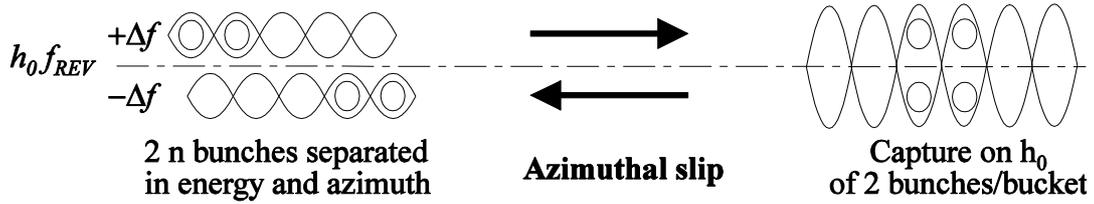

**Fig. 14:** Slip stacking

Although improvements can be introduced, like reducing the frequency difference towards the end of the process, the longitudinal contour enclosing a pair of bunches in the final bucket contains also a large area without particles. After filamentation, the macroscopic emittance is much more than doubled and longitudinal density is accordingly reduced.

## 5 Beam manipulations with broadband RF systems

For the needs of acceleration, high voltages are generally necessary and hence high impedance / high Q cavities are used to minimize the required RF power. Because of the limited bandwidth of these cavities, their field is a continuous sine-wave varying much more slowly than the revolution frequency. In specific cases, however, for example in storage rings, the required voltages can be small enough that a low cavity-impedance is acceptable. The available bandwidth can then allow for getting a voltage that departs completely from a continuous sine-wave.

### 5.1 Barrier/isolated bucket with single sine-wave

A single sine-wave pulsing at the revolution frequency of the beam generates an isolated or a barrier bucket depending upon its polarity and the sign of $\eta$ (Fig. 15).

In the case of the isolated bucket there is a stable ('synchronous') particle at the central zero-crossing of the sine-wave. Particles inside the sine-wave period can be captured and execute closed trajectories around it. Particles outside this bucket move along the full circumference.

In the case of the barrier bucket, the central zero-crossing of the sine-wave is an unstable position. The stable region is limited by the other zero-crossings and extends over all the circumference except the sine-wave.

Such a voltage can be obtained from a wideband resonator driven by a high power amplifier or from a limited bandwidth resonator driven by a large current generator.

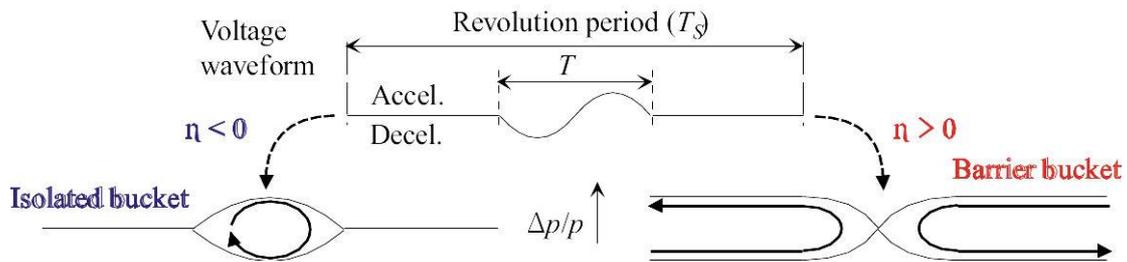

**Fig. 15:** Isolated/suppressed bucket

Beam dynamics is governed by the equations derived in Section 2.2. An isolated bucket is useful to capture a single bunch of small emittance in the debunched beam stack of an accumulator [8]. Barrier buckets are also typically used for high-intensity accumulation, to preserve gaps without beam and permit lossless beam transfers [15].

## 5.2 Barrier buckets with voltage pulses

A pair of voltage pulses with opposite polarities can also be used to generate a barrier or an isolated bucket (Fig. 16). Beam time structure and energy spread can be changed by modulating the amplitude and timing of the pulses as a function of time. These changes can be adiabatic provided that these modulations are slow enough. As a typical example, bunch compression is illustrated in Fig. 17.

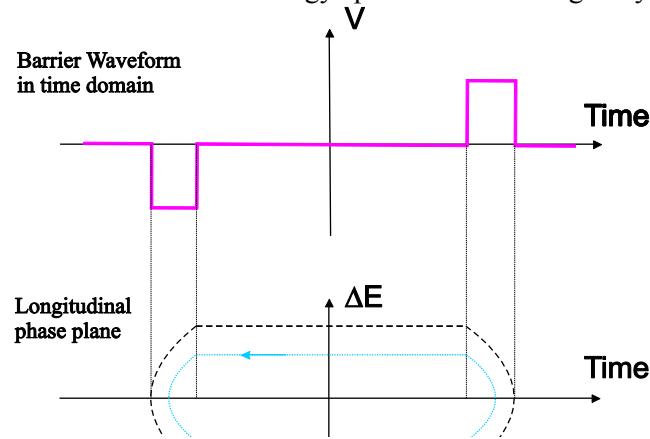

**Fig. 16:** Barrier bucket with voltage pulses

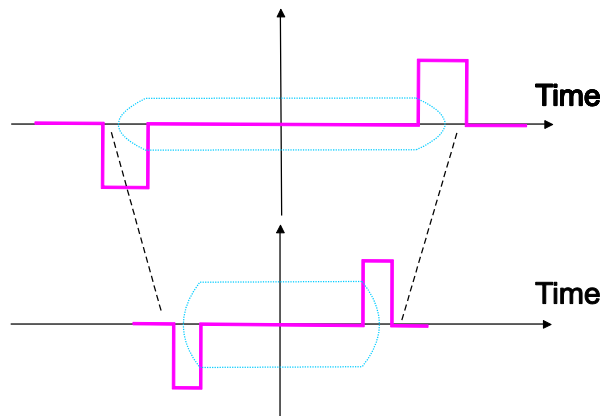

**Fig. 17:** Adiabatic bunch compression with voltage pulses

By adding more pulses and modulating them, sophisticated beam gymnastics can be done, similar to the ones feasible with conventional RF systems, but with the added flexibility resulting from the intrinsically fast time-response of the pulse generators [16].

# 6   Debunched beam gymnastics

**Phase displacement acceleration**

To keep the beam debunched, RF must be turned on without disturbing the longitudinal motion of the particles, and hence with a frequency which is outside of the beam spectrum. Shifting the RF frequency slowly towards and across the beam, the debunched beam can then be accelerated (or decelerated) by the passage of the empty RF buckets [17]. This is due to emittance preservation for the empty volume captured by the RF buckets. The resulting change of the stack mean energy is given by

$$\Delta E_{Stack} = A_{bucket} \frac{h}{T} = A_{bucket} f_{RF} \ . \tag{22}$$

A small voltage and a limited frequency range (a few per cent) are sufficient, and a large beam current and emittance can be handled. Repeating the process a large number of times, a significant energy change can be obtained. However, the acceleration/deceleration rate is small and the stack tends to degrade progressively as the number of traversals increases.

# 7   Practical implementation

The possible implementation and the effective performance of RF gymnastics in synchrotrons are constrained by a number of practical limitations. Apart from the basic hardware capabilities (number of simultaneous frequencies, minimum controllable voltage, etc.), the following must also be mentioned:

- maximum duration at constant field in the dipoles. This may force fast and non-adiabatic techniques or a degraded adiabaticity to be used.
- beam stability. The quality and reproducibility of performance of the final beam depends on the reproducibility of the initial conditions and the absence of collective beam instabilities during the process.
- control of the RF parameters. The proper operation of servo-loops (beam phase loop, radial or synchronization loop) all along the gymnastics is often critical for performance, and the unavoidable transients must be minimized, with their delayed effect quickly damped. Moreover, for good performance at high beam intensity, the beam loading in the RF cavities must be minimized so that local RF feedback and 'one-turn delay feedback' are often necessary.
- stability of the integrated dipole field during the gymnastics. This can be affected by drift or ripple and also by orbit bumps before beam ejection.

Setting-up time can be minimized by a preliminary analysis of the likely disturbances and the direct implementation of adequate corrective measures.